\documentclass{ws-ijqi}

\newcommand{\ket}[1]{\mbox{$\left|#1\right\rangle$}}

\begin{document}

\markboth{Jiannis K. Pachos}
{Three-spin interactions and entanglement in optical lattices}

%
\catchline{}{}{}{}{}
%

\title{Three-spin interactions and entanglement in optical lattices}

\author{JIANNIS K. PACHOS}

\address{Department of Applied Mathematics and Theoretical
Physics,\\ University of Cambridge,
Cambridge CB3 0WA, UK\\
j.pachos@damtp.cam.ac.uk}

\maketitle

\begin{history}
\end{history}

\begin{abstract}
The entanglement properties of some novel quantum systems are studied that
are inspired by recent developments in cold-atom technology. A triangular
optical lattice of two atomic species can be employed to generate a variety
of spin-$1/2$ Hamiltonians including effective three-spin interactions. A
variety of one or two dimensional systems can thus be realized that possess
multi-degenerate ground states or non-vanishing chirality. The properties of
these ground states and their phase transitions are probed with appropriate
measures such as the entropic entanglement and the spin chirality.
\end{abstract}


\section{Introduction}

Since the development of optical lattice
technology~\cite{Raithel,Mandel1,Mandel3}, considerable attention has been
paid to the experimental simulation of a variety of condensed matter
systems, such as spin chains~\cite{Jaks98,Kukl,Jask03,Duan}, and to the
realization of quantum computation~\cite{Deutsch}. Optical lattices allow to
probe and realize complex quantum models with unique properties in the
laboratory. Examples of particular interest in various areas of physics, are
systems with many-body interactions. The latter have been hard to realize
experimentally in the past due to the difficulty in controlling them
externally or isolating them from the environment~\cite{Mizel}. In contrast,
optical lattices give the means to realize such interactions. Their long
coherence times and their large degree of controllability that have been
studied theoretically and experimentally~\cite{Cirac1,Carl,Roberts} pave the
way towards the observation of ``higher order'' phenomena. Their
applications is of interest to cold atom technology as well as to condensed
matter physics and quantum information.

As a concrete physical system we shall consider the case of two different
atomic species, corresponding to different internal states of the atom,
superposed by optical lattices. The latter are standing laser waves
produced, for example, by a laser radiation trapped in an optical cavity.
Each atom can be considered as a dipole that experiences the standing wave
as a periodic sinusoidal potential. By employing two standing laser
radiations which differ in polarization or frequency it is possible to trap
the two different internal states of the atom and manipulate them
individually. For cold enough atoms and for sufficiently large amplitudes of
the standing waves it is possible to bring the system to the Mott insulator
phase where there is only one atom per site. This results in a two
dimensional Hilbert space for each site spanned by the two possible states
of an atom, namely $\ket{\uparrow}$ and $\ket{\downarrow}$, which
effectively represent a spin-1/2 system.

\begin{center}
\begin{figure}[!h]
\resizebox{!}{2.9cm} {\includegraphics{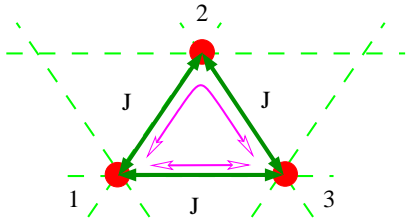} }
\caption{\label{triangle} The basic building block for the triangular
    lattice configuration. Three-spin
    interaction terms appear between sites $1$, $2$ and $3$. For
    example, tunneling between $1$ and $3$ can happen through two
    different paths, directly and through site $2$. The latter
    results in an exchange interaction between $1$ and $3$ that is
    influenced by the state of site $2$.}
\end{figure}
\end{center}

Even if the number of atoms at each site is constant throughout the lattice
there are still quantum effects present. They are mediated by virtual
transitions dictated by the tunnelling from one site to its neighbor sites
with coupling, $J$, and by collisions that take place when two or more atoms
are located within the same site with  coupling, $U$. The physical
conditions of weak tunnelling couplings, $J\ll U$, and weak atomic densities
assure that we are always in the Mott insulator regime with one atom per
site. To analyze the evolution of the system we apply perturbation theory
that restricts to the subspace of low energy states with one atom per site
which are virtually interacting with the rest energetically unfavorable
states. Considering up to the third order in perturbation theory it is shown
that a triangular configuration of the lattice, as in Figure~\ref{triangle},
results in a Hamiltonian that includes a variety of three-spin interactions.
They arise from the possibility of atoms tunnelling along two different
paths.

Several applications spring out from our studies. The systematic description
of the low energy Hamiltonian provides the means for the advanced control of
the three-spin interactions simulated in the lattice. Hence, different
physical models can be realized, with ground states that present a rich
structure such as multiple degeneracies and a variety of quantum phase
transitions \cite{Sachdev,Sachdev1,Pachos04}. Subsequently, these phases may
also be viewed as possible phases of the initial atomic system, that is in
the Mott insulator, where the behavior of the ground state can be controlled
at will. These multi-particle interactions can be realized, in principle,
with the near future technology.

\section{Effective three-spin interactions}
\label{eff}

\begin{center}
\begin{figure}[ht]
\resizebox{!}{2.5cm}
{\includegraphics{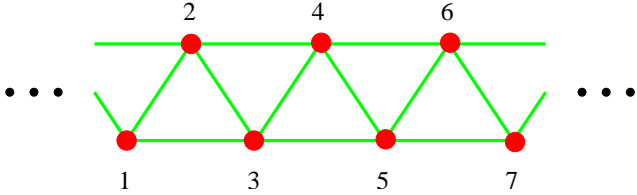} }
\caption{\label{chain} The one dimensional chain constructed out of
    equilateral triangles. Each triangle experiences the three-spin
    interactions presented in the previous.}
\end{figure}
\end{center}

Consider the low energy evolution of the triangular system given in Figure
\ref{triangle} of three atoms in three sites of the lattice. By applying
perturbation theory up to the third order we obtain the effective evolution
of the system. As we restrict to the low energy space of states given by
$\ket{\uparrow}$ and $\ket{\downarrow}$ for each site it is possible to
express the effective Hamiltonian in terms of Pauli spin-1/2 operators
explicitly given in the following
\begin{eqnarray}
H_{eff} =\sum_{j=1}^3 \Big[&&  \mathbf{B}_j\cdot
\mbox{\boldmath $\sigma$}_j + \lambda^{(1)}_j
\sigma^z_j \sigma^z_{j+1} + \lambda^{(2)}_j (\sigma^x_{j} \sigma^x_{j+1}
+\sigma^y_{j} \sigma^y_{j+1})
\nonumber \\ &&+\lambda^{(3)} \sigma^z_{j} \sigma^z_{j+1}
  \sigma^z_{j+2}+\lambda^{(4)}_j (\sigma^x_{j} \sigma^z_{j+1}
  \sigma^x_{j+2}+ \sigma^y_{j} \sigma^z_{j+1} \sigma^y_{j+2})
  \Big].
  \label{ham1}
\end{eqnarray}
Up to the second order the Hamiltonian includes a Zeeman field, an Ising
interaction and an $XX$ interaction. The three-spin interactions presented
in the last line can be viewed as the two spin interactions controlled by
the third spin through an additional $\sigma^z$ operator (see
Figure~\ref{triangle}). The couplings, $\mathbf{B}$ and $\lambda^{(i)}$, can
be given as expansions in ${J^\sigma}/U_{\sigma\sigma'}$ where
$\sigma=\uparrow,\downarrow$. These interactions can be straightforwardly
extended to the case of a semi-one dimensional model (see Figure
\ref{chain}) or a two dimensional one.

\begin{center}
\begin{figure}[!htb]
\resizebox{!}{6.0cm}{\includegraphics{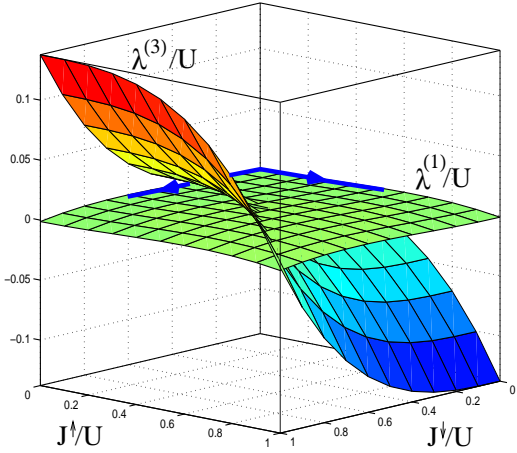}}
\caption{\label{zzz} The effective couplings $\lambda^{(1)}$ and
  $\lambda^{(3)}$
  are plotted against $J^{\uparrow}/U$ and $J^{\downarrow}/U$ for $U_{
  \uparrow \uparrow }=U_{\downarrow \downarrow }=2.12U$ and
  $U_{\uparrow \downarrow }=U$. The coupling $\lambda^{(1)}$ appears almost
  constant and zero as the unequal collisional terms can create a
  plateau area for a small range of the tunnelling couplings, while
  $\lambda^{(3)}$ can be varied freely to positive or negative values.}
\end{figure}
\end{center}

One can isolate different parts of the Hamiltonian (\ref{ham1}), each one
including a three-spin interaction term, by varying the tunnelling and/or
the collisional couplings appropriately so that particular terms in
considering up to the third order of the perturbation theory vanish, while
others are freely varied. In particular, we would like to see if we can make
the three-spin interaction dominating the two spin ones. Indeed, this is the
case for particular values of $J^\sigma$ and $U_{\sigma \sigma'}$ as we can
see in Figure
\ref{zzz}.

A particular example of a model that can be simulated in the optical lattice
is described by the one dimensional Hamiltonian of the form
\begin{equation}
H(B_x,B_z)= - \sum_j \big(B_x \sigma_j^x +B_z \sigma_j^z +
\sigma_j^z \sigma_{j+1}^z \sigma_{j+2}^z\big) \nonumber
\end{equation}
where all of its couplings can be arbitrarily and independently varied. The
three-spin interaction term of this Hamiltonian possesses fourfold
degeneracy in its ground state, spanned by the states $\{|\uparrow \uparrow
\uparrow\rangle,|\uparrow \downarrow \downarrow\rangle,|\downarrow
\uparrow \downarrow\rangle,|\downarrow \downarrow \uparrow\rangle
\}$. The criticality behavior of this model has been extensively
studied in the past \cite{Pens88,Chris}, where it is shown to present first
and second order phase transitions. In particular, for $B_z=0$ its self-dual
character can be demonstrated
\cite{Turb,Pens82}. To explicitly show that let us define the dual
operators
\begin{equation}
\bar \sigma^x_j \equiv \sigma^z_j \sigma^z_{j+1} \sigma^z_{j+2}
,\,\,\,\, \bar \sigma^z_j \equiv \prod_{k =0} ^\infty
\sigma^x_{i-3k} \sigma^x_{i-3k-1}, \nonumber
\end{equation}
that also satisfy the usual Pauli spin-1/2 algebra. We can re-express the
Hamiltonian $H(B_x,0)$ with respect to the new operators obtaining finally
\begin{equation}
H(B_x,0)= B_xH(B_x^{-1},0). \nonumber
\end{equation}
This equation of self-duality indicates that if there is one critical point
then it should be at $|B_x|= 1$.

\begin{center}
\begin{figure}[!ht]
\includegraphics[width=0.5\textwidth]{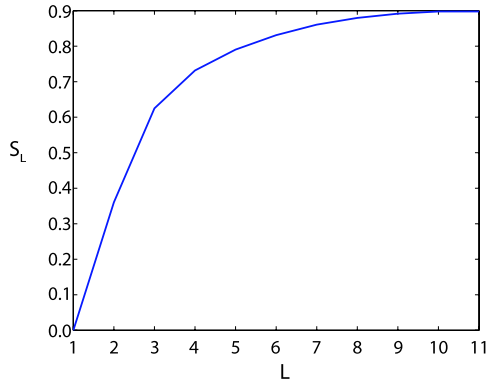}
\caption{\label{fig:XZZZ} Entropy of entanglement, $S_L$, as a function of
$L$ for a total of 19 spins. The plot shows a logarithmic behavior that
indicates criticality for $B_x=1$.}
\end{figure}
\end{center}

As it is not possible to explicitly solve the model we can employ numerical
techniques to establish its criticality. For that, we shall consider an
appropriate measure of entanglement, the entropy of
entanglement~\cite{Latorre}, that successfully reveals the criticality
behavior of our system. This measure is given by the von Neumann entropy,
$S_L$, of the reduced density matrix, $\rho_L$, of a subsystem of $L$
successive spins analytically given by
\begin{equation}
S_L  \equiv  \textrm{tr} (\rho_L \log \rho_L)\,\,\,\,\,
\textrm{with}\,\,\,\,\,\,\,\,
\rho_L \equiv {\textrm{tr}}_{N-L} |\psi \rangle \langle\psi|
\end{equation}
where $|\psi \rangle$ is the total ground state. We know that for
non-critical chains $S_L$ should be saturated for large enough values of
$L$. Indeed, this behavior is observed from the simulations when $B_x\neq
1$. On the other hand, when the system experiences second order criticality
we expect $S_L \approx {c+\bar{c}\over 6} \log L$, where $c$ is the central
charge of the corresponding conformal field theory and $\bar{c}$ is its
complex conjugate. The central charge uniquely corresponds to the critical
exponents of the energy and of the correlation length, $z$ and $\nu$,
respectively \cite{Sachdev}. Indeed, at $B_x=1$ we obtain Figure
\ref{fig:XZZZ} that shows the expected logarithmic progression. By a
logarithmic fitting we can deduce that $c\approx 4/5$, revealing that our
model is in the same universality class as the three states Potts model with
critical exponents $z=1$ and $\nu=3/4$.

\section{Complex tunnelling and topological effects}\label{complex}

Consider the case where the tunnelling couplings in the setup described
above are complex. This can be achieved by employing the electric moment
${\bf d}_e$ of the neutral atoms and an external magnetic field gradient.
Without restrictions, we assume that the dipole is created out of two
charges $e$ and $-e$ at positions ${\bf x}$ and ${\bf x}+{\bf d}$,
respectively, where ${\bf d}_e \equiv e{\bf d}$. Hence, its interaction with
the field can be described by the minimal coupling
\begin{eqnarray}
{\bf p} && \rightarrow {\bf p} +e\big[{\bf A}({\bf x}+{\bf d}) -{\bf A}({\bf
x})\big]
\nonumber \\ &&
\approx{\bf p} +e\big[{\bf A}({\bf x}) +
({\bf d}\cdot\mbox{\boldmath $\nabla$}){\bf A}({\bf x})-{\bf A}({\bf
x})\big]
={\bf p} + ({\bf d}_e\cdot \mbox{\boldmath $\nabla$}) {\bf A}({\bf x})
\end{eqnarray}
where the limit of small $|{\bf d}|$ with respect to the variation of the
magnetic field has been taken.

One can evaluate the phase produced by an electric dipole making a circular
path in the presence of a magnetic field. Indeed, by employing Stokes's
theorem we have
\begin{equation}
\phi= \oint_C ({\bf d}_e \cdot \mbox{\boldmath $\nabla$} )
({\bf A}({\bf x})\cdot d{\bf l}) =\int\int_\Sigma ({\bf
d}_e\cdot\mbox{\boldmath $\nabla$}) ({\bf B}\cdot d{\bf S})
\end{equation}
where $d{\bf l}$ and $d{\bf S}$ are the line and surface elements,
respectively.
\begin{center}
\begin{figure}[ht]
\resizebox{!}{3 cm}
{\includegraphics{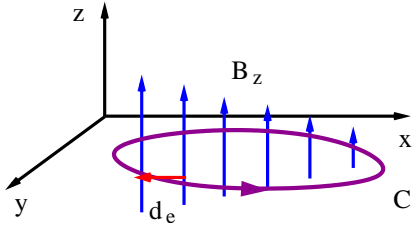} }
\caption{\label{magnetic} A dipole circulating a magnetic field that has a
constant gradient in the x-direction, while the field is homogenous in the
y-direction. The resulting quantum phase is equivalent to having a charged
particle circulating a homogeneous magnetic flux.}
\end{figure}
\end{center}
For example, consider the case of a magnetic field oriented along the
z-direction with a non-vanishing gradient along the x-axis, as in Figure
\ref{magnetic}. Then for an electric dipole oriented along x,
that performs a loop enclosing a surface $\Sigma$ on the x-y plane, we have
\begin{eqnarray}
\phi &&=
\int\int_\Sigma d_e {\partial B_z \over \partial x} dxdy
=\int \int_\Sigma q^*B^* dxdy
\end{eqnarray}
where the effective charge and magnetic field are given by
\begin{equation}
q^*B^*= d_e {\partial B_z / \partial x}.
\end{equation}
Hence, it is apparent that the phase obtained by a circulating dipole in the
presence of a constant gradient of the magnetic field can actually simulate
the effect of a circulating charge in a homogenous magnetic field. This
analogy can be generalized to the case of many (neutral) particles in a
superfluid phase such as a Bose-Einstein condensate. For the latter, an
alternative way to simulate the effect of a charged particle in the presence
of a magnetic field is given by just rotating it. Though, by employing
present atom chip technology the gradient of a magnetic field can achieve
$q^*B^*$ that is 10 to a 100 times stronger than the one obtained by the
rotating technique.

Consider now the case where the atoms are superposed by optical lattices. If
the dipole tunnels from one site to its neighboring one then there is a
phase factor contribution to the tunnelling couplings, $J=e^{i\phi}|J|$,
with
\begin{equation}
\phi=\int_{\mathbf{x}_i}^{\mathbf{x}_{i+1}}(\mathbf{d}_e \cdot
\mbox{\boldmath $\nabla$})\mathbf{A} \cdot d \mathbf{x}. \nonumber
\end{equation}
Here $\mathbf{x}_i$ and $\mathbf{x}_{i+1}$ denote the positions of the
lattice sites connected by the tunnelling coupling $J$. In order to isolate
the additional terms that appear in the case of complex tunnelling couplings
we should restrict ourselves to purely imaginary ones, i.e. we assume
$J_{j}^{\sigma}=
\pm i |J_{j}^{\sigma}|$. Then the effective Hamiltonian becomes
\begin{eqnarray}
H_{eff}=\sum_i \Big[&& {\bf B}\cdot {\mbox{\boldmath $\sigma$}}_i
+\tau^{(1)}\sigma^z_i\sigma^z_{i+1} +\tau^{(2)}(\sigma^x_i\sigma^x_{i+1}+
\sigma^y_i\sigma^y_{i+1}) \nonumber \\&&
+\tau^{(3)}(\sigma^x_i\sigma^y_{i+1}- \sigma^y_i\sigma^x_{i+1})+
\tau^{(4)}\epsilon_{lmn}\sigma^l_i \sigma^m_{i+1} \sigma^n_{i+2}
\Big],
\label{complexboson}
\end{eqnarray}
where $\epsilon_{lmn}$ with $\{l,m,n\}=\{x,y,z\}$ denotes the total
antisymmetric tensor in three dimensions and summation over the indices
$l,m,n$ is implied. The couplings $\tau^{(i)}$, appearing in
(\ref{complexboson}) are functions of the original tunnelling and
collisional couplings and can be varied at will. For example, by
appropriately tuning $J^a$ and $U_{ab}$ and with the aid of compensating
Zeeman terms we can set several of the $\tau^{(i)}$ couplings to zero
obtaining the Hamiltonian
\begin{equation}
H_{eff}=\sum_i \Big[\tau^{(1)} \mbox{\boldmath
$\sigma$}_i\cdot\mbox{\boldmath $\sigma$}_{i+1}+
\tau^{(4)}\mbox{\boldmath $\sigma$}_i\cdot(\mbox{\boldmath $\sigma$}_{i+1}
\times \mbox{\boldmath $\sigma$}_{i+2})\Big], \label{chiral}
\end{equation}
with $\mbox{\boldmath $\sigma$}=(\sigma^x,\sigma^y,\sigma^z)$. The
three-spin interaction term in (\ref{chiral}) is also known in the
literature as the chirality operator~\cite{Wen}. It breaks time reversal
symmetry of the system as a consequence of the externally applied field. It
is of great interest to study the ground state of this interaction in two
dimensions, where states with solitonic character~\cite{Wen,Rokhsar,Sen} are
expected to form the ground states.
\begin{center}
\begin{figure}[!ht]
\resizebox{!}{2.9cm} {\includegraphics{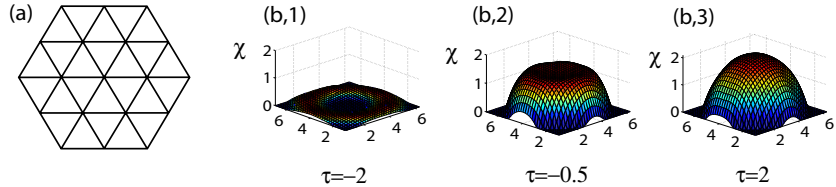} }
\caption{\label{xxyyboundaries} (a) The hexagonal structure with 19 spins
  on the vertices and the 24 triangles. (b) The chirality, $\chi$, as a
  function of the plaquette positioned on the plane for three different
  couplings $\tau$. The generation of a localized skyrmion is apparent
  for large and positive values of $\tau$.}
\end{figure}
\end{center}
\vspace{-0.5cm}

As a particular example, for studying the behavior of this three-spin
interaction term, we take a hexagonal configuration of 19 spin as in Figure
\ref{xxyyboundaries}(a). On this lattice we simulate Hamiltonian
(\ref{chiral}) for $\tau^{(1)}=\tau\cdot \tau^{(4)}$ with the additional
condition that the spins on the boundary experience a strong magnetic field
oriented in the z-direction. A numerical simulation has been performed to
obtain the chirality of the ground state, $\chi\equiv\langle\vec{\sigma}_i
\cdot\vec{\sigma}_j \times\vec{\sigma}_k \rangle$, for each triangular
plaquette of neighboring sites $i$, $j$ and $k$. In
Figure~\ref{xxyyboundaries}(b) the chirality for different values of $\tau$
on the plane of the hexagon is given for three distinctive values of $\tau$.

It is apparent that when $\tau$ is large and negative a ferromagnetic state
is dominant that has zero chirality. On the other hand, when $\tau$ is large
and positive frustration is dominating that cooperates with the chiral term
for the generation of a chiral ground state. This is clearly indicated in
Figure \ref{xxyyboundaries}(b) where non-zero chirality is localized in the
center of the hexagon indicating the presence of a skyrmion-like
configuration.

To better understand the nature of the ground state we look at the possible
expectation values the chirality operator can take for an arbitrary state of
the spins on a certain triangle. It is easy to verify~\cite{Simon} that a
product state gives $|\chi|\leq 1$, a bipartite entangled state gives
$|\chi|\leq 2$, while states with $|\chi|> 2$ are necessarily tripartite
entangled. From Figure~\ref{xxyyboundaries}(b) it is possible to see that
this is the case for $\tau \approx 2$ indicating that the ground state at
the center of the hexagon possesses tripartite entanglement.

\section{Conclusions}
\label{conclusions}

In this article we presented a variety of different spin interactions that
can be generated by a system of ultra-cold atoms superposed by optical
lattices and initiated in the Mott insulator phase. In particular, we have
been interested in the simulation and study of three-spin interactions
conveniently obtained in a lattice with equilateral triangular structure.
They appear by a perturbation expansion to the third order with respect to
the tunnelling transitions of the atoms when the dominant interactions are
collisions of atoms within the same site. Among the models presented here is
the $\sigma^z_i\sigma^z_{i+1} \sigma^z_{i+2}$ interaction as well as
interactions that explicitly break chiral symmetry. These models exhibit
degeneracy in their ground state and undergo a variety of quantum phase
transitions that can also be viewed as phases of the initial Mott insulator
\cite{Cruz}. The effect of these terms will eventually become significant
with the advance of experimental techniques.

The possibility to externally control most of the parameters of the
effective Hamiltonians at will qualifies cold atom technology as a unique
setup to study exotic systems such as chiral spin systems, fractional
quantum Hall systems or systems that exhibit high-$T_c$
superconductivity~\cite{Wen,Laughlin}. In addition, suitable applications
have been presented within the realm of quantum computation~\cite{Pachos03}
where three-qubit gates can be straightforwardly generated from the
three-spin interactions. Unique properties related to the criticality
behavior of the chain with three-spin cluster interactions have been
analyzed in~\cite{Pachos04} where the two-point correlations, used
traditionally to describe the criticality of a chain, seem to fail to
identify long quantum correlations, suitably expressed by particular
entanglement measures~\cite{Verstraete PC 03}. It is unadaptable then that
optical lattices offer a unique laboratory to simulate condensed matter
systems and to probe new properties of great theoretical and technological
interest.

\section{acknowledgements}

This work was supported by the Royal Society.


\begin{thebibliography}{0}

\bibitem{Raithel} A. Kastberg, W. D. Phillips, S. L. Rolston,
  R. J. C. Spreeuw, and P. S. Jessen, Phys. Rev. Lett. {\bf 74}, 1542
  (1995); G. Raithel, W. D. Phillips, and S. L. Rolston,
  Phys. Rev. Lett. {\bf 81}, 3615 (1998).

\bibitem{Mandel1}
M. Greiner, O. Mandel, T. Esslinger, T. W. H\"ansch, and I. Bloch, Nature
{\bf 415}, 39 (2002); M. Greiner, O. Mandel, T. W. H\"ansch, and I. Bloch,
Nature {\bf 419}, 51 (2002).

\bibitem{Mandel3}
O. Mandel, M. Greiner, A. Widera, T. Rom, T. W. H\"ansch, and I. Bloch,
Nature {\bf 425}, 937 (2003).

\bibitem{Jaks98} D. Jaksch, C. Bruder, J. I. Cirac, C.W. Gardiner,
and P. Zoller, Phys. Rev. Lett. {\bf 81}, 3108 (1998).

\bibitem{Kukl} A. B. Kuklov, and B. V. Svistunov,
Phys. Rev. Lett. {\bf 90}, 100401 (2003).

\bibitem{Jask03} D. Jaksch, and P. Zoller,
New Journal Phys. {\bf 5}, 56.1 (2003).

\bibitem{Duan} L. M. Duan, E. Demler, and M. D. Lukin,
Phys. Rev. Lett. {\bf 91}, 090402 (2003).

\bibitem{Deutsch}
I. H. Deutsch, G. K. Brennen, and P. S. Jessen, Forsch. der Phys. (2000).

\bibitem{Mizel} A. Mizel, and D. A. Lidar,
cond-mat/0302018.

\bibitem{Cirac1}
P. Rabl, A. J. Daley, P. O. Fedichev, J. I. Cirac, and P. Zoller,
cond-mat/0304026.

\bibitem{Carl}
S. E. Sklarz, I. Friedler, D. J. Tannor, Y. B. Band, and C. J. Williams,
Phys. Rev. A {\bf 66}, 053620 (2002).

\bibitem{Roberts}
D. C. Roberts, and K. Burnett, Phys. Rev. Lett. {\bf 90}, 150401 (2003).

\bibitem{Sachdev} S. Sachdev, {\em Quantum Phase Transitions},
Cambridge University Press (1999).

\bibitem{Sachdev1} P. Fendley, K. Sengupta, and S. Sachdev,
Phys. Rev. B {\bf 69}, 075106 (2004).

\bibitem{Pachos04}
J. K. Pachos, and M. B. Plenio, Phys. Rev. Lett. {\bf 93}, 056402 (2004).


\bibitem{Pens88} K. A. Penson, J. M. Debierre, and L. Turban,
Phys. Rev. B {\bf 37}, 7884 (1988).

\bibitem{Chris} J. Christian, A. d'Auriac, and F. Igl\'oi,
Phys. Rev. E {\bf 58}, 241 (1998).

\bibitem{Turb} L. Turban,
J. Phys. C: Solid State Phys. {\bf 15}, L65 (1982).

\bibitem{Pens82} K. A. Penson, R. Jullien, and P. Pfeuty,
Phys. Rev. B {\bf 26}, 6334 (1982).

\bibitem{Latorre}
J. I. Latorre, E. Rico, and G. Vidal, QIC {\bf 4}, 48 (2004).

\bibitem{Wen} X. G. Wen, F. Wilczek, and A. Zee, Phys. Rev. B {\bf 39},
  11413 (1989).

\bibitem{Rokhsar} D. S. Rokhsar, Phys. Rev. Lett. {\bf 65}, 1506
  (1990).
\bibitem{Sen} D. Sen, and R. Chitra, Phys. Rev. B {\bf 51}, 1922
  (1995).

\bibitem{Simon}
 G. A. Durkin, and C. Simon, quant-ph/0504072.

\bibitem{Cruz} C. D'Cruz, and J. K. Pachos, to arrear.

\bibitem{Laughlin} R. B. Laughlin, Science {\bf 242}, 525 (1988).

\bibitem{Pachos03} J. K. Pachos, and P. L. Knight,
Phys. Rev. Lett. {\bf 91}, 107902 (2003).

\bibitem{Verstraete PC 03} F. Verstraete, M. Popp, and J. I. Cirac,
Phys. Rev. Lett. {\bf 92}, 027901 (2004).



\end{thebibliography}
\end{document}